\documentclass[11pt]{article}
\usepackage{graphicx}
\usepackage{longtable,lscape}
\usepackage{amsfonts}
\begin{document}
\title{Classical Logical versus Quantum Conceptual Thought: Examples in Economics, Decision theory and Concept Theory}
\author{Diederik Aerts\\
        \normalsize\itshape
        Center Leo Apostel for Interdisciplinary Studies \\
         \normalsize\itshape
         Department of Mathematics and Department of Psychology\\
        \normalsize\itshape
        Vrije Universiteit Brussel, 1160 Brussels, 
       Belgium \\
        \normalsize
        E-Mail: \textsf{diraerts@vub.ac.be}
		\\ \\
		Bart D'Hooghe\\
        \normalsize\itshape
        Center Leo Apostel for Interdisciplinary Studies \\
         \normalsize\itshape
         Department of Mathematics \\
        \normalsize\itshape
        Vrije Universiteit Brussel, 1160 Brussels, 
       Belgium \\
        \normalsize
        E-Mail: \textsf{bdhooghe@vub.ac.be}
        }
\date{}
\maketitle              
\begin{abstract}
\noindent Inspired by a quantum mechanical formalism to model concepts and their
disjunctions and conjunctions, we put forward in this paper a specific
hypothesis. Namely that within human thought two superposed layers can be distinguished: (i)
a layer given form by an underlying classical deterministic process, incorporating essentially logical thought and its
indeterministic version modeled by classical probability theory; (ii) a
layer given form under influence of the totality of the surrounding conceptual landscape, where the different concepts figure as individual entities rather than (logical) combinations of others, with measurable quantities such as `typicality',
`membership', `representativeness', `similarity', `applicability',
`preference' or `utility' carrying the influences. We call the process in this second layer `quantum conceptual thought', which is
indeterministic in essence, and contains holistic aspects, but is equally well, although very differently,
organized than logical thought. A substantial part of the `quantum conceptual thought process' can be modeled by
quantum mechanical probabilistic and mathematical structures. We consider examples of three
specific domains of research where the effects of the presence of quantum conceptual thought and its deviations from classical logical thought have been noticed and
studied, i.e. economics, decision theory, and concept theories and which
provide experimental evidence for our hypothesis.
\end{abstract}


\section{Introduction}

We put forward in this paper a specific hypothesis. Namely that in human thought as a process two specifically structured and superposed layers can
be identified:

(i) A first layer that we call the `classical logical' layer. The thought process
within this layer is given form by an underlying classical logical conceptual process \cite{boole1854}. The manifest process itself may be, and generally will be, indeterministic,
but the indeterminism is due to a lack of knowledge about the underlying
deterministic classical process. For this reason the process within the
classical logical layer can be modeled by using a classical Kolmogorovian
probability description \cite{kolmogorov1956}, and eventually, in an idealized form, it could be
modeled as a stochastic process.

(ii) A second layer that we call the `quantum conceptual' layer. The thought process
within this layer is given form under influence of the totality of the surrounding conceptual landscape, where the different concepts figure as individual entities, also when they are combinations of other concepts, contrary to how this is the case in the classical logical layer where combinations of concepts figure as classical combinations of entities and not as individual entities. In this sense one can speak of a phenomenon of `conceptual emergence' taking place in this quantum conceptual layer, certainly so  for combinations of concepts. Quantum conceptual thought has been identified in different domains of knowledge and science related to
different, often as paradoxically conceived, problems in these domains. The sorts of measurable quantities being able to experimentally identify quantum conceptual thought have been different in these different domains, depending on which aspect of the conceptual
landscape was most obvious or most important for the identification of the deviation of classically expected values of these quantities. For
example, in a domain of cognitive science where representations of concepts
are studied, and hence where concepts and combinations of concepts, and
relations of items, exemplars, instances or features with concepts are
considered, measurable quantities such as `typicality', `membership',
`similarity' and `applicability' have been studied and used to experimentally put into evidence the deviation of what classically would be expected for the values of these quantities \cite{osherson1981,osherson1982,smith1984,hampton1988a,hampton1988b,storms1993,rips1995,storms1996,hampton1997a,storms1998,storms1999}. In decision theory measurable quantities such as `representativeness', `qualitative
likelihood' `similarity' and `resemblance' have played this role
\cite{kahneman1972,tversky1977,tversky1982,morier1984,wells1985,carlson1989,gavanski1991,bar-hillel1993,bagassi2007}. In a domain such as economics one has considered measurable quantities such as `preference', `utility' and
`presence of ambiguity' to put into evidence the deviation of classical values of these quantities \cite{allais1953,ellsberg1961}. The quantum conceptual thought process is indeterministic
in essence, i.e. there is not necessarily an underlying deterministic
process independent of the context. Hence, if analyzed deeper with the aim
of finding more deterministic sub processes, unavoidably effects of context
will come into play. Since all concepts of the interconnected web that forms the landscape of concepts and combinations of them attribute as individual entities to the influences reigning in this landscape, and more so since this happens dynamically in an environment where they are all quantum entangled structurally speaking, the nature of quantum conceptual thought contains aspects that we strongly identify as holistic and synthetic. However, the quantum conceptual thought process is not
unorganized or irrational. Quantum conceptual thought is as firmly structured as
classical logical thought but in a very different way. We believe that the reason why
science has hardly uncovered the structure of quantum conceptual thought is because
it has been believed to be intuitive, associative, irrational, etc...
meaning `rather unstructured'. As a consequence its structure has not been
sought for consistently since believed to be hardly existent. An important
second hypothesis that we put forward in this paper is that an idealized
version of this quantum conceptual thought process, or a substantial part of it, can be modeled as a quantum
mechanical process. To indicate this idealization or this part we have called it `quantum conceptual thought'. Hence we believe that important aspects of the basic structure of quantum conceptual thought can be uncovered as a
consequence of this quantum structure modeling.

The distinction of two modes of thought has been proposed by many and in many different ways. Already Sigmund Freud in his seminal work `The interpretation of dreams' made the proposal of considering thought as consisting of two processes, which he called primary and secondary \cite{freud1899}, a distinction that became popularized as conscious and subconscious. Somewhat later William James introduced the idea of `two legs of thought' where he specified one as `conceptual', being exclusive, static, classical and following the rules of logic, and the other one as `perceptual', being intuitive and penetrating. He expressed the opinion that `just as we need two legs to walk, we also need both conceptual and perceptual modes to think' \cite{james1910}. Remark that James used the connotation `conceptual' to indicate the classical logical mode, contrary to us using `conceptual' mainly with respect to the quantum structured mode. 
 Jean Piaget, in his study of child thought, introduced `directed or intelligent thought' which is conscious and follows the rules of logic and `autistic thought' which is subconscious and not adapted to reality in the sense that it creates a dream world \cite{piaget1923}. More recently, Jerome Bruner introduced the `paradigmatic mode of thought', transcending particularities to achieve systematic categorical cognition where propositions are linked by logical operators, and the `narrative mode thought', engaging in sequential, action-oriented, detail-driven thought, where thinking takes the form of stories and `gripping drama' \cite{bruner1990}. One of the authors of the present article has investigated aspects of different modes of thought and the influence of their presence on human cognitive evolution \cite{gaboraaerts2009}.
 
There are some fundamental differences between earlier versions in the literature of different modes of thought and the hypothesis about different layers of thought put forward in the present article. First of all, it is the specific mathematical structure of the quantum model, elaborated by one of the authors for the description of the combination of concepts in \cite{aerts2007a,aerts2007b,aerts2009}, that defines the structural aspects of the two layers of thought that we put forward in this article and how they are intertwined. This means that the nature of this double layered structure follows from a mathematical model for experimental data on the non classical aspects of thought identified in these different domains. Secondly, and directly related to the first difference, the double structure that we mention is a complex quantum entanglement, technically meaning a `superposition of two modes of thought' rather than an individual or separated or eventually parallel existence. In \cite{aerts2007a,aerts2007b,aerts2009} this entangled structure is analyzed in great detail, and it is shown that the presence of the two layers and the specific way they are entangled follows from the quantum field nature of the model developed in \cite{aerts2007b,aerts2009}. Also in the following of the present article we put forward some of these details. We believe that the superposed layers of thought have connections with the approaches of `two modes of thought' that have been considered in history \cite{freud1899,james1910,piaget1923,bruner1990,gaboraaerts2009}, and are planning to find out more about the details of such correspondences in future research. Actually we have worked mostly on the explanatory power for the specific examples in the different domains that we mentioned, and the rest of this article will focus on this.

The effects of the presence of quantum conceptual thought are observed in situations
where deviations of classical logical thought are apparent in a systematic and
intersubjective way, i.e. such that the effect can be measured and proven to
be not due to chance and be repeated quantitatively. In sections 2, 3 and 4
we give examples of three specific domains of research where the effects of
the presence of quantum conceptual thought and its deviations from classical logical thought
have been noticed and studied, i.e. economics, decision theory, and concept
theories. In section 5 we illustrate in detail the functioning of the two
modes of thought on the specific example of the `disjunction of concepts',
since it was indeed the quantum modeling of the disjunction of concepts in 
\cite{aerts2007a,aerts2007b,aerts2009} that made us propose the basic hypothesis of the presence
of the two superposed layers of classical logical thought and of quantum conceptual thought
within the human thought process as analyzed in the present article. There is a growing research activity in applying quantum structures 
to domains of science such as economics \cite{schaden2002,baaquie2004,haven2005,bagarello2006} and psychology and cognition \cite{aertsaerts1994,gaboraaerts2002,aertsgabora2005,busemeyerwangtownsend2006,busemeyermatthewwang2006,aerts2007a,aerts2007b,aerts2009} and language and artificial intelligence \cite{widdows2003,widdowspeters2003,aertsczachor2004,vanrijsbergen2004,aertsczachordhooghe2005,bruzacole2005}, and the study of the two layers of thought put forward in the present article is a contribution to this.

\section{Quantum Conceptual versus Classical Logical Thought in Economics}

Almost seven decades ago, John von Neumann and Oskar Morgenstern founded a
new branch of interdisciplinary research by applying game theory to
economics \cite{VN28,vn47}. Expected utility functions are used to model the
preferences of rational agents over different ventures with random
prospects, i.e. `betting preferences' over what can be called \textit{%
lotteries}. One of the basic principles of the von Neumann-Morgenstern
theory is Savage's Sure-Thing Principle (STP) \cite{savage1954}. Savage
introduced this principle in the following story: {\it A businessman contemplates
buying a certain piece of property. He considers the outcome of the next
presidential election relevant. So, to clarify the matter to himself, he
asks whether he would buy if he knew that the Democratic candidate were
going to win, and decides that he would. Similarly, he considers whether he
would buy if he knew that the Republican candidate were going to win, and
again finds that he would. Seeing that he would buy in either event, he
decides that he should buy, even though he does not know which event
obtains, or will obtain, as we would ordinarily say}. This assumption is the
independence axiom of expected utility theory: independence means that if a
person is indifferent between simple lotteries $L_{1}$ and $L_{2}$, the
agent is also indifferent between $L_{1}$ mixed with an arbitrary simple
lottery $L_{3}$ with probability $p$ and $L_{2}$ mixed with $L_{3}$ with the
same probability $p$.

Problems in economics such as Allais paradox \cite{allais1953} and Ellsberg
paradox \cite{ellsberg1961} show an inconsistency with the predictions of
the expected utility hypothesis, namely a violation of the STP. As an
illustration we consider the situation put forward by Daniel Ellsberg \cite%
{ellsberg1961} which was mostly meant to illustrate the so called `ambiguity
aversion': persons prefer `sure choices' over `choices that contain
ambiguity'. Consider the following hypothetical experiment. Imagine an urn
known to contain 30 red balls and 60 black and yellow balls, the latter in
unknown proportion. One ball is to be drawn at random from the urn. To `bet
on Red' will mean that you will receive a prize $a$ (say \$100) if you draw
a red ball (`if Red occurs') and a smaller amount $b$ (say, \$0) otherwise
(`if not-Red occurs'). One considers the following 4 actions: (I) `a bet on
red', (II) `a bet on black', (III) `a bet on red or yellow', (IV) `a bet on
black or yellow', and the pairs of gambles (I, II), (III, IV). Ellsberg
found that a very frequent pattern of response is I preferred to II, and IV
is preferred to III. Less frequent is: II preferred to I, and III preferred
to IV. Both of these violate the STP, which requires the ordering of I to II
to be preserved in III and IV (since the two pairs differ only in the
pay-off when a yellow ball is drawn, which is constant for each pair). The
first pattern, for example, implies that the subject prefers to bet `on' red
rather than `on' black; and he also prefers to bet `against' red rather than
`against' black. This contradiction indicates that preferences of subjects
are inconsistent with the independence axiom of expected-utility theory.

Approaches have been developed, such as for example the `info-gap approach',
where it is supposed that the considered person implicitly formulates
`info-gap models' for the subjectively uncertain probabilities. The person
then tries to satisfy the expected utility and to maximize the robustness
against uncertainty in the imprecise probabilities. This robust-satisfying
approach can be developed explicitly to show that the choices of
decision-makers should display precisely the preference reversal which
Ellsberg observed \cite{ben-haim2006}.

\section{Effects of Quantum Conceptual Thought in Decision Theory}

The situation considered in Ellsberg paradox shows a great similarity to
situations considered in decision theory with respect to what is called the
disjunction effect \cite{tversky1992}, where also STP is violated, and the
conjunction fallacy \cite{tversky1982,tversky1983}. The disjunction effect
occurs when decision makers prefer option $x$ (versus $y$) when knowing that
event $A$ occurs and also when knowing that event $A$ does not occur, but
they refuse $x$ (or prefer $y$) when not knowing whether or not $A$ occurs.
A well-known example of such disjunction effect is the so-called Hawaii
problem, in which following two situations are considered \cite{tversky1992}:

\noindent \textbf{1. Disjunctive version:}\newline
{\it Imagine that you have just taken a tough qualifying examination. It is the
end of the fall quarter, you feel tired and run-down, and you are not sure
that you passed the exam. In case you failed you have to take the exam again
in a couple of months---after the Christmas holidays. You now have an
opportunity to buy a very attractive 5-day Christmas vacation package to
Hawaii at an exceptionally low price. The special offer expires tomorrow,
while the exam grade will not be available until the following day. Would
you:\newline
$x$ buy the vacation package\newline
$y$ not buy the vacation package\newline
$z$ pay a \$5 non-refundable fee in order to retain the rights to buy the
vacation package at the same exceptional price the day after
tomorrow---after you find out whether or not you passed the exam}.

\noindent \textbf{2. Pass/fail version:\newline
} {\it Imagine that you have just taken a tough qualifying examination. It is the
end of the fall quarter, you feel tired and run-down, and you find out that
you passed the exam (failed the exam. You will have to take it again in a
couple of month---after the Christmas holidays). You now have an opportunity
to buy a very attractive 5-day Christmas vacation package to Hawaii at an
exceptionally low price. The special offer expires tomorrow. Would you:%
\newline
$x$ buy the vacation package\newline
$y$ not buy the vacation package\newline
$z$ pay a \$5 non-refundable fee in order to retain the rights to buy the
vacation package at the same exceptional price the day after tomorrow}.

In the Hawaii problem, more than half of the subjects who know the outcome
of the exam (54\% in the passed condition and 57\% in the fail condition)
choose option $x$ --- buy the vacation package, but only 32\% do it in the
uncertain condition of not knowing the outcome of the exam. This is a
crucial demonstration that Tversky and Shafir produced to show the
disjunction effect. Here decision makers prefer option $x$ (to buy the
vacation package) when they are in a certain condition (passed exam) and
they also prefer $x$ when they are not in that condition (failed exam), but
they refuse $x$ (or prefer $z$) when they don't know which condition they
are in (they don't know the outcome of the exam).

Next to the disjunction effect in decision theory an effect called the
conjunction fallacy was identified. This effect occurs when it is assumed
that specific conditions are more probable than a single general one. The
most oft-cited example of this fallacy originated with Amos Tversky and
Daniel Kahneman \cite{tversky1982,tversky1983}: {\it Linda is 31 years old,
single, outspoken, and very bright. She majored in philosophy. As a student,
she was deeply concerned with issues of discrimination and social justice,
and also participated in anti-nuclear demonstrations. Which is more
probable? (1) Linda is a bank teller. (2) Linda is a bank teller and is
active in the feminist movement}. 85\% of those asked chose option (2).
However, if classical Kolmogorovian probability theory is applied, the probability of two events occurring together
(in `conjunction') will always be less than or equal to the probability of
either one occurring alone.

\section{The Guppy Effect in Concept Theories}

Situations that can be compared with the foregoing described paradoxes,
effects, fallacies, in economics and decision theory have been studied in the
field of concepts representation. In \cite{osherson1981} the concepts 
\textit{Pet} and \textit{Fish} and their conjunction \textit{Pet-Fish} are
considered, and observed that an item such as \textit{Guppy} turns out to be
a very typical example of \textit{Pet-Fish}, while it is neither a very
typical example of \textit{Pet} nor of \textit{Fish}. Hence, the typicality
of a specific item with respect to the conjunction of concepts can behave in
an unexpected way. The problem is often referred to as the `pet-fish
problem' and the effect is usually called the `guppy effect'. The guppy
effect is abundant and appears almost in every situation where concepts are
combined \cite%
{osherson1981,osherson1982,smith1984,hampton1988a,hampton1988b,storms1993,rips1995,storms1996,hampton1997a,storms1998,storms1999}%
.

The guppy effect was not only identified for the typicality of items with
respect to concepts and their conjunctions but also for the membership
weights of items with respect to (i) concepts and their conjunction \cite%
{hampton1988a}, (ii) concepts and their disjunction \cite{hampton1988b}. For
example, the concepts \textit{Home Furnishings} and \textit{Furniture} and
their disjunction \textit{Home Furnishings or Furniture} was one of the pair
of studied concepts. With respect to this pair, Hampton considered the item 
\textit{Ashtray}. Subjects estimated the membership weight of \textit{Ashtray%
} for the concept \textit{Home Furnishings} to be 0.3 and the membership
weight of the item \textit{Ashtray} for the concept \textit{Furniture} to be
0.7. However, the membership weight of \textit{Ashtray} with respect to the
disjunction \textit{Home Furnishings or Furniture} was estimated only 0.25,
less than both weights with respect to both concepts apart. This means that
subjects found \textit{Ashtray} to be `less strongly a member of the
disjunction \textit{Home Furnishings or Furniture}' than they found it to be
a member of the concept \textit{Home Furnishings} alone or a member of the
concept \textit{Furniture} alone. If one thinks intuitively of the `logical'
meaning of a disjunction, this is an unexpected result. Indeed, someone who
finds that \textit{Ashtray} belongs to \textit{Home Furnishings}, would be
expected to also believe that \textit{Ashtray} belongs to \textit{Home
Furnishings or Furniture}. Equally so for someone who finds that \textit{%
Ashtray} is a piece of \textit{Furniture}.

This deviation from what one would expect of a standard classical
interpretation of the disjunction was called `underextension' in \cite%
{hampton1988b}. Although the item \textit{Ashtray} with respect to the
disjunction of the concepts \textit{Home Furnishings} and \textit{Furniture}
shows underextension, and hence the presence of the effect of
underextension, could be interpreted as due to `ambiguity aversion', also
the inverse effect occurs in many occasions. For example with respect to the
pair of concepts \textit{Fruits} and \textit{Vegetables} and their
disjunction \textit{Fruits or Vegetables}, for the item \textit{Olive} the
membership weights with respect to \textit{Fruits}, \textit{Vegetables} and 
\textit{Fruits or Vegetables} were respectively 0.5, 0.1 and 0.8. This means
that \textit{Olive} is estimated by subjects to belong `more' to \textit{%
Fruits or Vegetables} than to any one of the concepts apart.

The examples in the different domains of science, economics, decision theory
and concept theories, that we gave in the foregoing sections illustrate the
presence of the two layers of thought which we called the classical logical layer and
the quantum conceptual layer. In the next section we analyze the two layers of
thought in detail on a specific example from the domain of concept theory.



\section{Classical Logical and Quantum Conceptual Thought}

Let us analyze in detail the structure of the two layers of thought on the
specific example of the `disjunction of concepts'. It is indeed the quantum
modeling of the disjunction of concepts as elaborated in \cite{aerts2007a,aerts2007b,aerts2009}, and more specifically in \cite{aerts2009}, section 7,
that reveals the presence of the two superposed layers of classical logical thought
and of quantum conceptual thought for the situation of the disjunction of two
concepts. We will analyze the two layers in detail here now.

Consider the situation of a subject performing one of the experiments of
Hampton \cite{hampton1988a,hampton1988b}, and more concretely the subject is
estimating the membership weight of an item $X$ with respect to the
disjunction `$A$ or $B$' of the two concepts $A$ and $B$. To make the
situation even more concrete, consider as an example the item \textit{Apple}
with respect to the pair of concepts \textit{Fruits} and \textit{Vegetables }%
and their disjunction `\textit{Fruits or Vegetables}'. A subject might go
about more or less as follows: `An apple is certainly a fruit, but it is
definitely not a vegetable. But hence it is certainly also a `fruit or a
vegetable', since it is a fruit'. This `reasoning' fits into a classical
Boolean scheme \cite{boole1854}, indeed if one proposition is true then also the
disjunction of this proposition with another proposition is true. Of
course, in general the membership weight which is an average over the yes/no
attributions by the individual subjects will not be equal to $1$, as it is
the case for the items \textit{Apple}, but will be between $0$ and $1$. Consider for example the item \textit{Pumpkin} with respect to the same pair of
concepts. As has been measured in \cite{hampton1988b}, and also to be found
in Table 1 of \cite{aerts2009}, the membership weights of this item with
respect to the concepts \textit{Fruits}, \textit{Vegetables} and
their disjunction \textit{Fruits or Vegetables} are respectively 0.7, 0.8
and 0.925. We prove in \cite{aerts2009}, section 2.2, that with these
membership weights it is possible for a deterministic logical underlying
process to exist, such that these weights are the results of the classical
Kolmogorovian chance for a specific subject to choose `for' or `against'
membership of the item \textit{Pumpkin} with respect to the concepts \textit{%
Fruits}, \textit{Vegetables} and \textit{Fruits or Vegetables} respectively.
This means that for the thought process that enrolls when a subject is
deciding `for' or `against' membership of \textit{Pumpkin} with respect to
the pair of concepts \textit{Fruits}, \textit{Vegetables} and \textit{Fruits
or Vegetables} it is possible to find an underlying process that is
deterministic and enrolls following classical logic. The
manifest process measured in \cite{hampton1988b} can as a consequence be
modeled by means of a classical stochastic process, where the probability,
giving rise to the weights, is due to a lack of knowledge of an underlying
deterministic classical logic process.

Let us consider now a third item with respect to the same pair of concepts,
namely the item \textit{Olive}. In \cite{hampton1988b} the following weights
were measured, 0.5 with respect to \textit{Fruits}, 0.1 with respect to 
\textit{Vegetables} and 0.8 with respect to \textit{Fruits or Vegetables}.
Hence this is a case called overextension in \cite{hampton1988b}. We prove
in \cite{aerts2009}, section 2.2, that for these weights it is not possible
to find a Kolmogorovian representation. This means that these weights
cannot be obtained by supposing that subjects reasoned following classical
logic and that the weights are the result of a lack of knowledge about the
exact outcomes given by each of the individual subjects. Indeed, if 50\% of
the subjects has classified the item \textit{Olive} belonging to \textit{%
Fruits}, and 10\% has classified it as belonging to \textit{Vegetables} than
following a classical reasoning at most 60\% of the subjects (corresponding
with the set-theoretic union of two mutually distinct sets of subjects) can
classify it as belonging to \textit{Fruits or Vegetables}. Hence, this means that these
weights arise in a distinct way. Some individual subjects must have chosen {\it Olive} as a member of {\it Fruits or Vegetables} and `not as a member' of {\it Fruits} and also `not as a member' of {\it Vegetables}, otherwise the weights 0.5, 0.1 and 0.8 cannot result. It is here that the second layer of
thought, namely quantum conceptual thought, comes into play. Concretely this means
that for the item \textit{Olive} the subject does not reason in a logical
way, but rather directly wonders whether \textit{Olive} is a member or not a
member of \textit{Fruits or Vegetables}. In this quantum conceptual thought process
the subject considers \textit{Fruits or Vegetables} as a new concept and not
as a classical logical disjunction of the two concepts \textit{Fruits} and 
\textit{Vegetables} apart. Hence the subject gets influenced in his or her choice in favor or against membership of {\it Olive} with respect to {\it Fruits or Vegetables} by the presence of the new concept \textit{Fruits or Vegetables}. This is the reason why we say that within
the quantum conceptual thought process it is the emergence of a new concept, i.e. the concept {\it Fruits or Vegetables} within the landscape of existing concepts, i.e. {\it Fruits}, {\it Vegetables} and {\it Olive}, that gives rise to the deviation of the membership weight that would be expected following classical logic. And it is the probability to decide for or against membership that is influenced by the presence of this new concept within the landscape of existing concepts. Concretely, in this case, the subject estimates whether 
\textit{Olive} is characteristic for the new concept \textit{Fruits or
Vegetables}, hence whether \textit{Olive} is one of these items where indeed
one can doubt whether it is a \textit{Fruit} or a \textit{Vegetable}. And
right so, for \textit{Olive} this is typically the case, which is the reason
that its weight with respect to \textit{Fruits or Vegetables} is big, namely
0.8, as compared to rather small, 0.5 and 0.1 with respect to the concepts 
\textit{Fruits} and \textit{Vegetables} apart, and most important `bigger than the sum of both' (0.8 is strictly bigger than 0.5+0.1), which makes a classical explanation impossible, as we prove in \cite{aerts2009}. The proof in \cite{aerts2009}
that the weights for the item \textit{Olive} with respect to the pair of
concepts \textit{Fruits}, \textit{Vegetables} and \textit{Fruits or
Vegetables} cannot be modeled within a Kolmogorovian probability structure,
has as a consequence that there cannot exist an underlying deterministic
process giving rise to these weights. Hence, it means that the conceptual
thought process that takes place when a subject decides `for' or `against'
membership of the item \textit{Olive} with respect to the concepts \textit{%
Fruits}, \textit{Vegetables} and \textit{Fruits or Vegetables} is
intrinsically indeterministic. In \cite{aerts2009} we also show that the
conceptual thought process, hence the weights that it produces with respect
to the different conceptual structures, when different possible decisions are considered, can be modeled by means of a quantum
mechanical probability structure.

\section{Experimental Arguments for Quantum Conceptual Thought}

The disjunction effect, one of the effects where following our hypothesis
`quantum conceptual thought is present', can be tested in various experimental
settings. We will show in this section that specific situations that
have been investigated demonstrate the presence of quantum conceptual thought. More
concretely, the experiments that we will describe in this section show that what is crucial to explain the effects measured is; (i) that the whole and overall conceptual landscape is relevant for the situation
considered, and; (ii) that the concepts play a role as individual entity, hence different from what this role would be if they were just combinations, i.e. disjunctions or conjunctions, of existing concepts. This is the reason why these experiments show the presence of what
we have called `quantum conceptual thought' in the situations typical of the
disjunction effect. This also explains why the often proposed explanation of
`uncertainty aversion', although it can play a role, as long as it fits into
the conceptual structure, is not the main cause of the disjunction effect.
Let us clarify this concretely by referring to the examples of the foregoing
section: \textit{Olive} scores `higher' than classically expected with respect
to \textit{Fruits or Vegetables}, hence contrary to what the disjunction effect
would provoke if it were caused by `uncertainty aversion'. Hence, if one would reflect in terms of `uncertainty
aversion' as explanation of the disjunction effect, it would mean that 
\textit{Olive} `likes' uncertainty instead of disliking it. Of course, what
is really at work, as we explained in the foregoing section, is that \textit{%
Olive} is characteristic conceptually for items that are \textit{Fruits or
Vegetables}, hence \textit{Olive} indeed `likes' uncertainty in this sense
metaphorically speaking. In the experiments that we consider in the
following of this section, we will see proofs of the fact that it is the
overall conceptual landscape that is at the origin of the effects, and such that `concepts influence as  individual entities and not as classical logical combinations of other existing concepts' which is
exactly what we have called the `presence of quantum conceptual thought'.

In \cite{bar-hillel1993} Maya Bar-Hillel and Efrat Neter explored the
possibility of extending the conjunction fallacy \cite{tversky1983} to a
more general extension fallacy, using natural disjunctive categories and
including problems that involve no compound events, which allowed to check
whether the fallacy results from incorrect combination rules. Students
received brief case descriptions and ordered 7 categories according to the criteria: (a) probability of membership, (b) willingness to bet on
membership. Let us give some examples to illustrate the type of test made. A detailed description of a person is given: {\it Danielle, sensitive and introspective. In high
school she wrote poetry secretly.
Did her military service as a
teacher. Though beautiful, she
has little social life, since she
prefers to spend her time reading
quietly at home rather than
partying}. The question is: What does she study? And then the alternatives to choose from are: {\it Literature}; {\it Humanities}; {\it Physics} or {\it Natural Sciences}. The second person considered is: {\it Oded: Did his military service as a
combat pilot. Was a brilliant
high school student, whose
teachers predicted for him an
academic career. Independent
and original, diligent and honest.
His hobbies are shortwave radio
and Astronomy}. The question is again: What does he
study? The alternatives to choose from are again: {\it Physics}; {\it Natural Sciences}; {\it Literature} or {\it Humanities}.

One of the basic rules of classical probability is violated in all cases tested in \cite{bar-hillel1993}. Let us point out more in detail what happens. Consider the {\it Danielle} case. Following the rules of classical logic, studying {\it Literature} implies studying {\it Humanities}. This means that it is `always more probable that {\it Danielle} studies {\it Humanities} than that {\it Danielle} studies {\it Literature}'. However, 82\% of the subjects indicated {\it Literature} and not {\it Humanities} as more probable, and 75\% of the subject preferred to bet on {\it Literature} instead of {\it Humanities}. This effect is called a `disjunction fallacy' by the authors. The disjunction fallacy turned out much less pronounced for the less probable choices. Concretely, 45\% of the subjects indicated {\it Physics} and not {\it Natural Sciences} as more probable for {\it Danielle}, and 27\% of the subjects preferred to bet on {\it Physics} instead of {\it Natural Sciences}. A very similar and even more pronounced result for the second example of {\it Oded}. Subjects indicated {\it Physics} and not {\it Natural Sciences} as more probable, and preferred to bet on {\it Physics} instead of {\it Natural Sciences}. For the less probable choices again the effect was less. Other but similar situations were tested and all revealed the same disjunction fallacy. The results support the view that the disjunction fallacy in probability judgments is due to representativeness, i.e. the degree of
correspondence between an instance and a concept. Concepts that are `more representative' are preferred, even if such a choice violates the rules of classical probability and logic. More specifically, {\it Literature} is preferred to {\it Humanities} for {\it Danielle} and {\it Physics} to {\it Natural Sciences} for {\it Oded} because as concepts they represent better their respective personalities. On the contrary, {\it Natural Sciences} is preferred to {\it Physics} for {\it Danielle} and {\it Humanities} is preferred to {\it Literature} for {\it Oded} because as concepts both represent less badly their respective personalities. The fact that this inversion of the effect takes place for `badly representing concept' shows that the effect is not due to some kind of overall preference for basic concepts, which {\it Literature} and {\it Physics} are, as compared to superordinate concepts, which {\it Humanities} and {\it Natural Sciences} are. Since stimuli used in the experiments could not be judged by combination
rules, these results also go against claims that probability fallacies `stem primarily from the incorrect rules people use to
combine probabilities \cite{gavanski1991}'.
We have mentioned earlier `representativeness' as one of the measurable quantities that reveals the presence of what we have called quantum conceptual thought, and this is what we see at play here. The experiments in \cite{bar-hillel1993} show that a concept such as {\it Humanities}, although originally conceived as the disjunction of {\it Languages}, {\it Literature}, {\it History}, {\it Philosophy}, {\it Religion}, {\it Visual and Performing Arts} and {\it Music}, `does not behave as the classical logical disjunction of these basic concepts', exactly analogously with the non classical logical behavior we have put forward in detail in the foregoing section of the present article for the disjunction {\it Fruit or Vegetables} of the two concepts {\it Fruits} and {\it Vegetables}. Equally so for the concept {\it Natural Sciences}, although originally being the disjunction of {\it Astronomy}, {\it Physics}, {\it Chemistry}, {\it Biology} and {\it Earth Sciences}, it does not behave as a concept in the way it should when simply being the classical logical disjunction of the different natural sciences. The quantum mechanical approach elaborated in \cite{aerts2007a,aerts2007b,aerts2009} models this non classical logical behavior.

Another set of experiments related to the disjunction effect was organized by Maria Bagassi and Laura Macchi 
\cite{bagassi2007}. Their aim was to show that the disjunction effect does not depend
on the presence of uncertainty (pass or fail the exam) but on the
introduction into the text-problem of a non-relevant goal. This indicates in a very explicit way that it is the overall conceptual
landscape that gives form to the
disjunction effect. More specifically Bagassi and Macchi point out that, option $z$ (`pay a \$5
non-refundable fee in order to retain the rights to buy the vacation package
at the same exceptional price the day after tomorrow---after you find out
whether or not you passed the exam') contains an
unnecessary goal, i.e. that one needs to `pay to
know', which is independent of the uncertainty condition.
In this sense, their hypothesis is that the choice of option $z$ occurs as a consequence of the
construction of the discourse-problem itself (\cite{bagassi2007}, p. 44).
Four experiments were performed in which various modifications with respect
to option $z$ were considered, ranging from (1) eliminating from the text and
option $z$ any connection between the `knowledge of the outcome' and the `decision'; (2) eliminating option $z$, limiting the
decision to $x$ (`buy') or $y$
(`not buy'); (3) making option $z$ more
attractive; (4) render the procrastination option $z$ more onerous. The
experimental results support the view that the disjunction effect does not
depend on the uncertainty condition itself, but on the insertion of the
misleading goal `paying to know' in the
text-problem. Eliminating it (but maintaining the uncertainty condition) the
disjunction effect vanishes (exp. 1 and 2). Also, if choice for $z$ is
sensible (exp. 3), most subjects choose it. If option $z$ is onerous (exp.
4), it is substantially ignored. In this sense, option $z$ is not a real
alternative to $x$ and $y$, but becomes an additional premise that conveys
information, which changes the decisional context. Hence the crucial factor
is the relevance of the discourse-problem of which $z$ is one element,
rather than certainty versus uncertainty. These results support the view
that the disjunction effect can be realized by applying a \textit{suitable}
decisional context rather than an \textit{uncertain} decisional context. If the suitable decision context, or in our terminology `specific conceptual landscape surrounding the decision situation', is what lies at the origin of the disjunction
effect, then this shows that what we have called `quantum conceptual thought' is taking
place during the process that gives rise to the disjunction effect. Following the experimental results of \cite{bagassi2007} one can argument that `the specific conceptual landscape surrounding the decision situation' plays a principal role in shaping the disjunction effect.

\section{Conclusions}

Inspired by a quantum mechanical formalism to model concepts and their
disjunctions and conjunctions, we put forward in this paper a specific
hypothesis. Namely that two superposed layers exist within human thought:

(i)
A layer which we call the `classical logical layer', and which is given form by an underlying classical deterministic process, giving rise to classical logical
thought and its indeterministic manifestation modeled by classical
probability theory. We refer to thought in this layer as `classical logical thought'.

(ii) A layer which we call the `quantum conceptual layer', and which is given form by the influence and structure of the overall conceptual landscape, where concepts and also all combinations of concepts exercise their influence on an individual basis. This means that combinations of concepts emerge in this layer as new individual concepts and not just logical combinations which is what they are in the classical logical layer. In this quantum conceptual layer the global and holistic effects of the overall conceptual landscape can be experimentally detected by measuring specific quantities that may be different depending on the domain under consideration, and a substantial part of this layer can be modeled by quantum mechanical
probabilistic structures. The effects of the presence of conceptual thought can be observed in
situations where deviations of logical thought are apparent in a systematic
and intersubjective way, i.e. such that the effect can be measured and
proven to be not due to chance and be repeated quantitatively. We have
considered examples of three specific domains of research where the effects
of the presence of conceptual thought and its deviations from logical
thought have been noticed and studied, i.e. economics, decision theory, and
concept theories. In concept theories quantities such as `typicality',
`membership', `similarity' and `applicability' can be measured and shown to deviate from what their values should be if thought would be classical logical. In decision
theory this role is played by quantities such as `representativeness', `qualitative likelihood'
`similarity' `resemblance', and in economics
quantities like `preference', `utility' and `presence of ambiguity' put into evidence the presence of quantum conceptual thought by deviating from their classical logical values. We have illustrated in detail the
functioning of the two layers of thought on the specific example of the
`disjunction of concepts', and we analyze two experimental investigations on the
disjunction effect that put into evidence the presence of quantum conceptual thought
with respect to this disjunction effect.


\end{document}